\documentclass[sigconf]{acmart}

\AtBeginDocument{%
  }
    
\usepackage{graphicx}
\usepackage{subcaption}
\usepackage{enumitem}
\usepackage{booktabs}
\usepackage{array}
\usepackage{adjustbox}
\usepackage{mathtools}
\usepackage{cellspace}
\usepackage{float}
\usepackage{multirow}
\usepackage{multibib}
\usepackage[ruled,vlined]{algorithm2e}
\usepackage{cancel}
\usepackage{xcolor}
\usepackage{amsmath}

\usepackage{amssymb}

\begin{document}

\title{Mining Intraday Risk Factors via Hierarchical Reinforcement Learning with Transferred Options}

\author{Wenyan Xu}
\affiliation{%
  \institution{School of Statistics and Mathematics, Central University of Finance and Economics}
  \city{Beijing}
  \country{China}}
\email{2022211032@email.cufe.edu.cn}

\author{Jiayu Chen}
\affiliation{%
  \institution{Industrial Engineering, Purdue University}
  \city{West Lafayette, IN}
  \country{USA}}
\email{chen3686@purdue.edu}

\author{Dawei Xiang}
\affiliation{%
  \institution{Dept. of Computer Science and Engineering, University of Connecticut}
  \city{Storrs, CT}
  \country{USA}}
\email{ieb24002@uconn.edu}

\author{Chen Li}
\affiliation{%
  \institution{Computer Network Information Center, Chinese Academy of Sciences}
  \city{Beijing}
  \country{China}}
\email{lichen@sccas.cn}

\author{Yonghong Hu}
\affiliation{%
  \institution{School of Statistics and Mathematics, Central University of Finance and Economics}
  \city{Beijing}
  \country{China}}
\email{huyonghong@cufe.edu.cn}

\author{Zhonghua Lu}
\affiliation{%
  \institution{Computer Network Information Center, Chinese Academy of Sciences}
  \city{Beijing}
  \country{China}}
\email{zhlu@sccas.cn}



\begin{abstract}
Traditional risk factors like beta and momentum often lag behind fast-moving markets in capturing stock return volatility, while statistical methods such as PCA and factor analysis struggle with nonlinear patterns. Genetic programming (GP) can uncover nonlinear structures but tends to produce overly complex formulas, and Transformer-based approaches lack built-in mechanisms for evaluating factor quality. To address these gaps, we propose an end-to-end reinforcement learning framework based on Hierarchical Proximal Policy Optimization (HPPO), unifying factor generation and evaluation. HPPO uses two hierarchical PPO models: a high-level policy that learns feature weights and a low-level policy that composes operators. Factor effectiveness is directly optimized using the Pearson correlation between the generated factors and target volatility as the reward. We further introduce Transferred Options (TO), enabling rapid adaptation by pretraining on historical data and fine-tuning on recent data. Experiments show HPPO-TO outperforms baselines by 25\% across major HFT markets.
\end{abstract}





\maketitle

\section{Introduction}

Risk factors are crucial for investors, translating historical trading data into forward-looking measures of return volatility that inform risk identification and decision-making. Traditionally, these factors—such as beta, size/value, momentum, and liquidity—are hand-crafted by domain experts. However, manual construction has significant drawbacks, including debates over factor selection, weak correlations to realized volatility, and an inability to adapt quickly to shifting markets.

Statistical approaches like principal component analysis \cite{de2021factor} and factor analysis\cite{harman1976modern} help uncover latent risk factors, but their linear frameworks cannot capture complex nonlinear relationships in the data. Deep risk models (DRMs)\cite{lin2021deep}  address this limitation by leveraging deep neural networks to learn implicit factors that better model return volatility, enhancing covariance estimation in Markowitz-style mean-variance frameworks. Nevertheless, these learned embeddings are often opaque and lack interpretability, limiting their practical utility for investment professionals who need transparent, actionable insights.
To bridge this gap, interpretable risk factors in closed-form mathematical expressions remain essential. 

Genetic programming (GP)\cite{lin2019revisiting,zhang2020autoalpha,cui2021alphaevolve} is widely used for this purpose, casting risk factor discovery as a symbolic regression (SR) problem\cite{cranmer2020discovering}. In quantitative factor mining, SR leverages historical trading data to uncover precise mathematical patterns. GP constructs binary expression trees from historical data, enabling the discovery of complex, nonlinear patterns without manual feature engineering. However, GP frequently suffers from “bloat”,\cite{fitzgerald2013bootstrapping} producing overly complex formulas as it prioritizes fitness without effectively constraining expression size. While the Transformer-based method\cite{xu2025hrft} generates more concise expressions, they still lack intrinsic mechanisms for evaluating factor quality, highlighting the ongoing need for interpretable, high-quality risk factors in quantitative finance.

To automatically evaluate generated risk factors, we leverage a reinforcement learning framework that uses reward signals to directly guide factor quality. To address the complexity of factor mining, we introduce Hierarchical Proximal Policy Optimization (HPPO), which breaks the process into two sub-tasks: a high-level policy that selects feature weights, and a low-level policy that composes factors with mathematical operators (e.g., $\mathit{log}, \mathit{tan}, \mathit{*}, \mathit{/}$). Using the Pearson correlation with realized volatility as the reward, HPPO seamlessly unifies factor generation and evaluation.

\begin{figure}[b]
\centering
\includegraphics[width=0.5\textwidth] 
{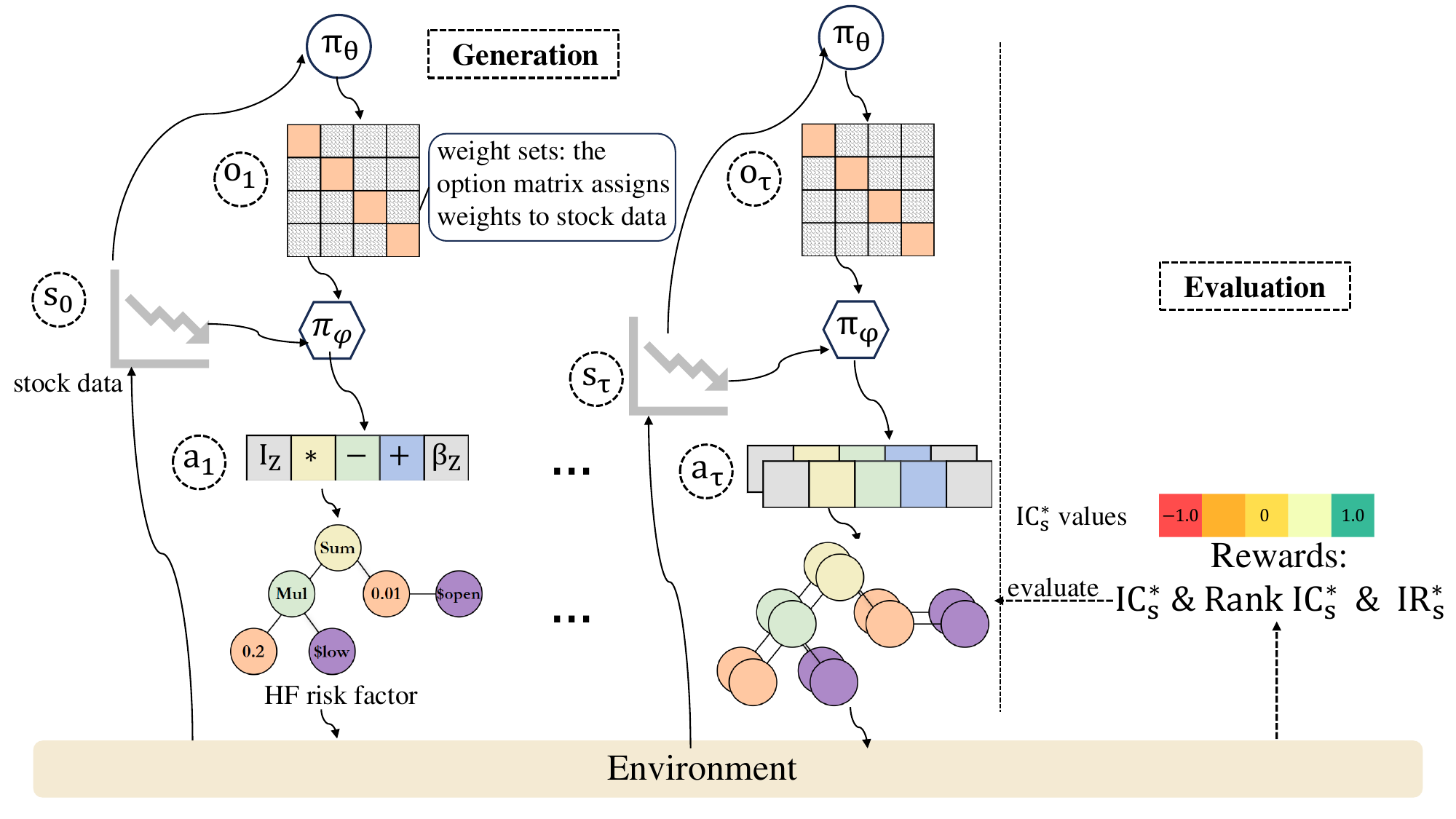}
\caption{ At each time step, the high-level policy selects a weight set (option) for the raw stock features, represented by four one-hot vectors. This option matrix, which also serves as the key/value matrix for the multi-head attention (MHA) mechanism ($d_k = d_v$), embeds each option. The low-level policy then composes stock features using operators such as ``$+$'' and ``$\cos()$''. The effectiveness of the resulting high-frequency risk factor is evaluated using modified $IC^*$, Rank $IC^*$, and $IR^*$ as rewards.}
\label{fig:hppo-to-framework.pdf} 
\end{figure}

Building upon HPPO, we develop HPPO with Transferred Options (HPPO-TO), integrating transfer learning. In HPPO-TO, the high-level policy is pre-trained on historical high-frequency trading (HFT) data and fine-tuned on recent data, significantly cutting training time and computational demands. The high-level policy generates "options"—weight combinations—while the low-level policy continuously refines operator combinations based on rewards, extracting robust features from both historical and current data. This combination enhances both adaptability and transferability.

We benchmark HPPO-TO against two genetic programming methods \cite{kommenda2020parameter}, one deep learning method \cite{petersen2019deep}, and two hierarchical reinforcement methods—Double Actor-Critic (DAC) \cite{zhang2019dac} and baseline HPPO \cite{chen2023multi}—across U.S. (S\&P 500), Chinese (HS 300) HFT markets. Extensive experiments confirm HPPO-TO generates superior high-frequency risk factors, consistently outperforming competitors in realistic portfolio simulations.

Our contributions include:
\begin{itemize}[noitemsep,topsep=0pt]
\item Introducing an end-to-end automated approach for high-frequency risk factor generation with concise mathematical forms, validated via portfolio optimization and short-term risk tasks.
\item Developed HPPO-TO, an efficient, transfer learning-enhanced method for faster and more accurate factor adaptation.
\item Demonstrating HPPO-TO’s superior performance, achieving approximately 25\% excess returns in diverse international HFT markets.
\end{itemize}

\section{Hierarchical Reinforcement Learning based on Transfer Options}
Hierarchical Reinforcement Learning, grounded in the option framework \cite{sutton1999between}, addresses complex tasks by decomposing them into subtasks at multiple levels of abstraction \cite{ghavamzadeh2006hierarchical,hengst2012hierarchical,li2017efficient,pateria2021hierarchical, xiang2025promptsculptor}. In the context of factor mining, the high-level policy assigns weights to stock features (e.g., high/low prices, trading volume), while the low-level policy composes these features using mathematical operators (e.g., $\mathit{+, -, *, /}$) (see Figure \ref{fig:hppo-to-framework.pdf}). In our framework, raw stock inputs—\textit{open}, \textit{close}, \textit{high}, \textit{low}, \textit{volume}, and \textit{vwap}—are transformed into concise, formula-based high-frequency risk factors. This hierarchical structure enables efficient identification of informative features and effective operator combinations, allowing HRL to scale to large state-action spaces more effectively than conventional reinforcement learning methods \cite{collins2018learning,li2019hierarchical,hutsebaut2022hierarchical}. Furthermore, HRL’s modular design supports transfer learning, enabling sub-policies trained in one environment to be adapted to new trading scenarios \cite{chen2023multi,eppe2022intelligent,xu2025finmultitime,xu2025learning}, thereby improving generalization and adaptability.

\subsection{High-level policy} \label{High-level policy}

The high-level policy inputs high-frequency stock features $X_{t+1} = \{ x_{t+1}^1,\, \ldots,\, \\ x_{t+1}^5 \}$ at the current time step and the weight combination (the option) $Z_t = \{ z_t^1,\, \ldots,\, z_t^5 \}$ from the previous time step, then outputs updated weights $Z_{t+1} = \{ z_{t+1}^1,\, \ldots,\, z_{t+1}^5 \}$ to align high-frequency risk factors with the target.

\noindent\textbf{State Space} comprises the stock features $X_{t+1}$ at the current time step and the weight combination $Z_t$ from the previous time step.

\noindent\textbf{Option Space} refers to the weight vector $W_{t+1} = \{ w_{t+1}^1, \ldots, w_{t+1}^5 \}$ at the current time step, which reallocating feature importance and guides the model in identifying their predictive significance.

\noindent\textbf{Reward.} Factor quality is gauged with three statistics:  
(i) the cross‑sectional Pearson correlation (IC) between factor values $E_i(f)$ and next‑day realised volatility (RV) $y_i$\,\cite{cohen2009pearson};  
(ii) the Spearman correlation (Rank‑IC)\,\cite{hauke2011comparison}; and  
(iii) the information ratio (IR), i.e.\ the mean IC divided by its standard deviation.  
\vspace{-0.3cm}
\begin{align}
\text{IC}_t         &= \mathbb{E}_t\!\left[\sigma(E_i(f),y_i)\right],\\
\text{RankIC}_t     &= \mathbb{E}_t\!\left[\sigma\bigl(r(E_i(f)),r(y_i)\bigr)\right],\\
\text{IR}_t         &= \frac{\overline{\text{IC}_t}}{\operatorname{std}(\text{IC}_t)},
\end{align}
where $\sigma$ is the Pearson kernel and $r(\cdot)$ denotes ranks.
\vspace{-0.5cm}

\subsection{Low‑level Policy}

Conditioned on $X_{t+1}$ and $Z_{t+1}$, the low‑level policy constructs an analytic risk factor by selecting an operator sequence $A_{t+1}$.  

\noindent\textbf{State space.} Its state is $(X_{t+1},Z_{t+1})$.

\noindent\textbf{Action space.} The action space consists of four binary operators—\texttt{add}, \texttt{sub}, \texttt{mul}, \texttt{div}—and ten unary operators:  
\texttt{inv}, \texttt{sqr}, \texttt{sqrt}, \texttt{sin}, \texttt{cos}, \texttt{tan}, \texttt{atan}, \texttt{log}, \texttt{exp}, and \texttt{abs}.  
These unary and binary operators are randomly combined to form the operator sequence $A_{t+1}$, which is recursively applied to $(x_{t+1}^k, z_{t+1}^k)$ to generate closed-form factors.

\noindent\textbf{Reward.} The low‑ and high‑level policies share the same reward signal, ensuring coherent optimisation across the hierarchy.


\begin{algorithm}[t]
    \caption{HPPO-TO: Risk Factor Generator}
    \label{alg:algorithm}
    \DontPrintSemicolon 
    \KwIn{Pre-trained low-level policy $\pi_\phi$, weight embedding matrix $\boldsymbol{W}_C$, initial features $S_0$, initial weights $Z_0$}
    \KwOut{Optimized risk factors}
    
    Initialize $\pi_\phi$ with pre-trained options\;
    Initialize $\boldsymbol{W}_C$\;
    Set $S_t \leftarrow S_0$\;
    Set $Z_t \leftarrow Z_0$\;
    
    \While{not converged}{
        \For{$t = 1$ to $\text{RolloutLength} - 1$}{
            Embed weights: $Z_t \leftarrow \boldsymbol{W}_C^T Z_t$\;
            Sample next weights: $Z_{t+1} \sim \pi_\theta(Z_{t+1} \mid S_t, Z_t)$\;
            Embed $Z_{t+1} \leftarrow \boldsymbol{W}_C^T Z_{t+1}$\;
            Sample operator sequence: $A_t \sim \pi_\phi(A_t \mid S_t, Z_{t+1})$\;
            Compute baselines: $b^{\text{high}}(S_t, Z_t)$, $b^{\text{low}}(S_t, Z_{t+1})$\;
            Apply $A_t$ to $S_t$; observe $S_{t+1}$ and reward $IC^*$\;
        }
        
        \For{$t = \text{RolloutLength}$ to $1$}{
            Option advantage: $Adv_t^Z = Ret_t - b^{\text{high}}(S_{t-1}, Z_{t-1})$\;
            Operator advantage: $Adv_t^A = Ret_t - b^{\text{low}}(S_{t-1}, Z_t)$\;
        }
        
        \While{$i < \text{PPO Optimization Epochs}$}{
            Update $\theta \leftarrow \text{PPO}(\frac{\partial \pi_\theta}{\partial \theta}, Adv^Z)$\;
            Update $\phi \leftarrow \text{PPO}(\frac{\partial \pi_\phi }{\partial \phi}, Adv^A)$\;
        }
    }
\end{algorithm}

\begin{table}[b]
\centering
\caption{Main results of HS300 Index and S\&P500 Index. "(x)" represents the standard deviation of $IC^*$, $Rank IC^*$ and $IR^*$, and the rest are the mean values. "↑" indicates that the larger the value, the better (\textit{Bold} indicates the optimal values).}
\label{tab:comparison_methods}
 \resizebox{0.5\textwidth}{!}{%
\begin{tabular}{ccccccc}
\toprule
\multirow{2}{*}{\textbf{Method}} & \multicolumn{3}{c}{\textbf{S\&P500}} & \multicolumn{3}{c}{\textbf{HS300}} \\
\cmidrule(r){2-4} \cmidrule(l){5-7}
& \textbf{IC*↑} & \textbf{Rank IC*↑} & \textbf{IR*↑} & \textbf{IC*↑} & \textbf{Rank IC*↑} & \textbf{IR*↑} \\
\midrule
\multirow{2}{*}{\centering \textbf{DSR}} 
& 0.0437 & 0.0336 & 0.2707 & 0.0391 & 0.0456 & 0.4021 \\ 
& (0.0054) & (0.0045) & (0.0353) & (0.0064) & (0.0063) & (0.0362)  \\
\multirow{2}{*}{\centering \textbf{HRFT}} 
& 0.0662 & 0.0720 & 0.4960 & 0.0618 & 0.0683 & \textbf{0.6460} \\ 
& (0.0077) & (0.0085) & (0.0817) & (0.0083) & (0.0088) & (0.1002)  \\
\multirow{2}{*}{\centering \textbf{GPLEARN}} 
& 0.0388 & 0.0437 & 0.4876 & 0.0494 & 0.0480 & 0.3599 \\ 
& (0.0068) & (0.0085) & (0.0307) & (0.0062) & (0.0063) & (0.0368)  \\
\multirow{2}{*}{\centering \textbf{GENEPRO}} 
& 0.0470 & 0.0549 & 0.2098 & 0.0444 & 0.0460 & 0.4257 \\ 
& (0.0067) & (0.0163) & (0.0339) & (0.0049) & (0.0075) & (0.0442)  \\
\multirow{2}{*}{\centering \textbf{DAC}} 
& 0.0421 & 0.0386 & 0.3461 & 0.0465 & 0.0508 & 0.3729 \\ 
& (0.0054) & (0.0041) & (0.0387) & (0.0078) & (0.0064) & (0.0454)  \\
\multirow{2}{*}{\centering \textbf{HPPO}} 
& 0.0569 & 0.0597 & 0.3642 & 0.0511 & 0.0557 & 0.4644 \\ 
& (0.0057) & (0.0057) & (0.0060) & (0.0043) & (0.0059) & (0.0364)  \\
\multirow{2}{*}{\textbf{Ours*}} 
& \textbf{0.0719} & \textbf{0.0774} & \textbf{0.7266} & \textbf{0.0739} & \textbf{0.0766} & 0.5680 \\
& (0.0072) & (0.0054) & (0.0448) & (0.0058) & (0.0059) & (0.0412) \\
\bottomrule
\end{tabular}
}
\end{table}


\subsection{Overall Framework}
The overall objective function of HPPO-TO is defined as
\begin{equation}
L = \mathbb{E}{\theta, \phi} \left[ \sum{t=1}^T r(S_t, A_t) \right].
\end{equation}
By computing gradients with respect to $\theta$ and $\phi$, we derive the actor-critic structure:
\begin{equation}
\begin{aligned}
\nabla_{\theta}L &= \mathbb{E} \left[ \sum_{t=1}^{T} \nabla_{\theta} \log\pi_{\theta}(Z_t|S_{t-1}, Z_{t-1}) \big(Ret_t - b^{high}(S_{t-1}, Z_{t-1})\big) \right], \\
\nabla_{\phi}L &= \mathbb{E} \left[ \sum_{t=1}^{T} \nabla_{\phi} \log\pi_{\phi}(A_{t-1}|S_{t-1}, Z_t) \big(Ret_t - b^{low}(S_{t-1}, Z_t)\big) \right].
\end{aligned}
\end{equation}
Here, $Ret_t$ is the return at time $t$, and $b^{high}$, $b^{low}$ are baselines (critics) for the high- and low-level policies, respectively. The advantage functions are $Ret_t - b^{high}(S_{t-1}, Z_{t-1})$ and $Ret_t - b^{low}(S_{t-1}, Z_t)$. Both policies $\pi_\theta$ and $\pi_\phi$ are optimized using PPO \cite{schulman2017proximal}.  Notably, $b^{high}$ can be parameterized using $b^{low}$:
\begin{equation}
b^{high}(S_{t-1}, Z_{t-1}) = \sum_{Z_t} \pi_\theta (Z_t | S_{t-1}, Z_{t-1}) b^{low}(S_{t-1}, Z_t).
\end{equation}
In finance, data distributions frequently shift over time. HPPO-TO addresses this by first pre-training on large-scale historical HFT data, then fine-tuning on recent data to stay aligned with current market conditions and avoid model obsolescence. A key advantage of HPPO-TO is its use of transferred options: options learned in similar historical contexts are directly applied to current data, eliminating the need for costly retraining. This continual transfer of knowledge enables HPPO-TO to adapt efficiently and achieve superior results compared to standard HRL methods. Full implementation details are provided in Algorithm~\ref{alg:algorithm}.
\begin{table}[b]
\centering
\caption{Information of stock data used in the experiments}
\label{tab:stock_data}
\scalebox{0.95}{
\begin{tabular}{cc|c}
\toprule
& \textbf{U.S. Market} & \textbf{Chinese Market} \\
& \textbf{S\&P500 (1min)} & \textbf{HS300 (1min)} \\
\midrule
\textbf{Pre-train} & 2023/01/03-2023/08/31 & 2022/10/31-2023/06/31 \\ 
\textbf{Train} & 2023/08/31-2023/12/29 & 2023/06/31-2023/10/31 \\ 
\textbf{Sample Size} & 18,330,000 & 7,964,160 \\ 
\bottomrule
\end{tabular}
}
\end{table}

\begin{figure}[b]
    \centering
    \begin{subfigure}[b]{0.98\linewidth}
        \centering
        \includegraphics[width=\linewidth]{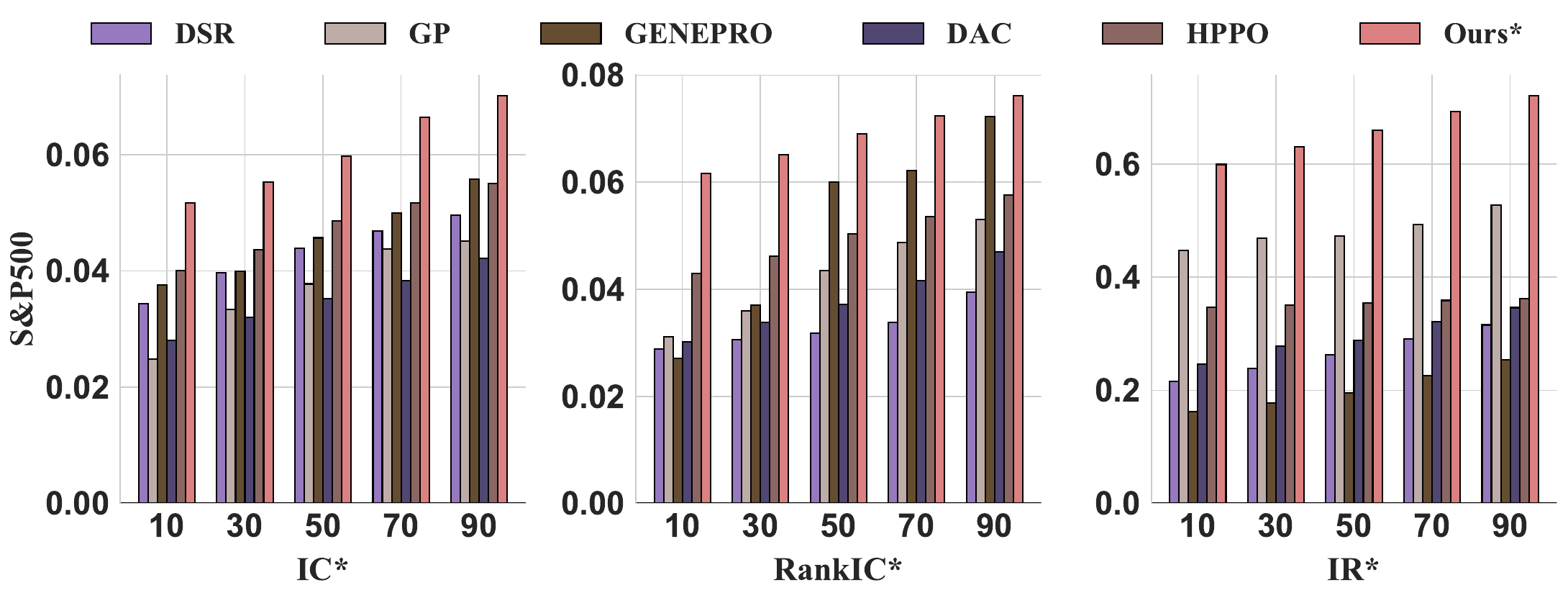}
        \caption{S\&P500}
    \end{subfigure}
    \vspace{0.2cm}
    \begin{subfigure}[b]{0.98\linewidth}
        \centering
        \includegraphics[width=\linewidth]{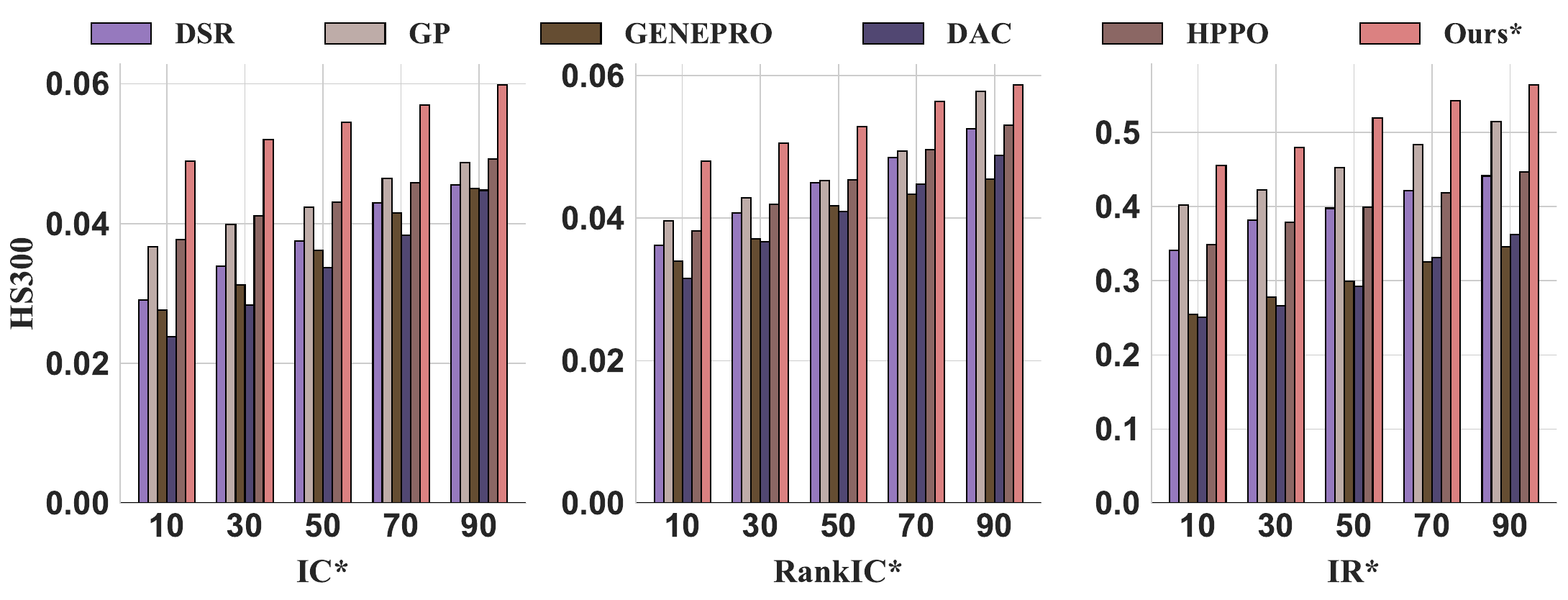}
        \caption{HS300}
    \end{subfigure}
    \caption{Performance comparison of all methods for generating factors with different factor pool sizes in terms of $IC^*$, $Rank IC^*$ and $IR^*$ metrics. The x-axis denotes the size of the factor pool, corresponding to the number of risk factors, and the y-axis indicates the metric values for the factors.}
    \label{fig:1}
\end{figure}

\begin{table}[b]
  \centering
  \caption{Top 5 risk factor expressions based on $IC^*$ values in the factor collection (S\&P500 Index).}
  \label{tab:10hf_risk}
  \resizebox{0.5\textwidth}{!}{
    \begin{tabular}{lccc}
      \toprule
      \textbf{No.} & \textbf{high-frequency risk factor} & \textbf{Option Index} & \textbf{IC*} \\
      \midrule
      \textbf{1} & $(0.1\cdot\text{open})\cdot(0.3\cdot\text{low})-(0.18\cdot\text{volume})/(0.4\cdot\text{vwap})$ & 2 & 0.0854 \\
      \textbf{2} & $(0.1\cdot\text{open})-(0.1\cdot\text{low})\cdot(0.5\cdot\text{high})\cdot(0.2\cdot\text{close})$ & 5 & 0.0587 \\
      \textbf{3} & $(0.3\cdot\text{open})\cdot(0.09\cdot\text{low})^{(0.3\cdot\text{high})}-(0.1\cdot\text{close})$ & 1 & 0.0567 \\
      \textbf{4} & $(0.18\cdot\text{volume})^{(0.4\cdot\text{vwap})}$ & 2 & 0.0541 \\
      \textbf{5} & $(0.1\cdot\text{open})/(0.3\cdot\text{low})$ & 1 & 0.0464 \\
      \bottomrule
    \end{tabular}
  }
\end{table}
\vspace{-0.2cm}

\section{Experiments}

\subsection{Experiment Settings}

We introduce datasets, baselines, and evaluation metrics.

\noindent\textbf{Data \& Evaluation Metrics}\quad Inputs are $m$-dimensional raw trading data \textit{(open/low/high/close/volume/vwap)} $X \in \mathbb{R}^m$ from constituents of HS300\footnote{https://www.wind.com.cn/} and S\&P500 indices (see Table \ref{tab:stock_data}). The target is one-day-ahead RV, defined as:
\begin{equation}
{RV}(t, j; n) = \sum_{j=1}^n (\ln P_{t, j} - \ln P_{t, j-1})^2
\end{equation}
where $n$ is intraday intervals and $P_{i, j}$ is the closing price for day $i$, interval $j$. Trading periods vary by market: China (240 mins) and U.S. (390 mins). Thus, intervals are: $M^{HS300}=240$, $M^{S\&P500}=390$. The dataset splits into historical (Pre-train) and current (Train) datasets, totaling approximately 26 million samples. Factor quality evaluation employs three standard positive metrics, incentivizing model performance.

\noindent\textbf{Baselines}\quad Our proposed method is evaluated against two HRL and three SR benchmarks:
\begin{itemize}
\item \textbf{DL-based}: \textbf{DSR} employs recurrent neural networks with risk-seeking policy gradients for factor generation \cite{petersen2019deep}. \textbf{HRFT} treats mathematical expression generation as a language problem, leveraging transformer models end-to-end \cite{xu2025hrft}.
\item \textbf{GP-based}: \textbf{GPLEARN}\footnote{https://github.com/trevorstephens/gplearn}, specialized for SR tasks compatible with scikit-learn, and \textbf{GENEPRO}\footnote{https://github.com/marcovirgolin/genepro}, supporting broader input types through tree-based structures.
\item \textbf{HRL-based}: \textbf{HPPO} simultaneously trains high-level policy $\pi_\theta$ and low-level policy $\pi_\phi$ with PPO advantage functions \cite{chen2023multi}. \textbf{DAC} integrates two parallel actor-critic structures within an options framework, utilizing state value functions \cite{zhang2019dac}.
\end{itemize}

\begin{figure}[b]
  \centering
  \begin{subfigure}{0.48\linewidth}
    \includegraphics[width=\linewidth]{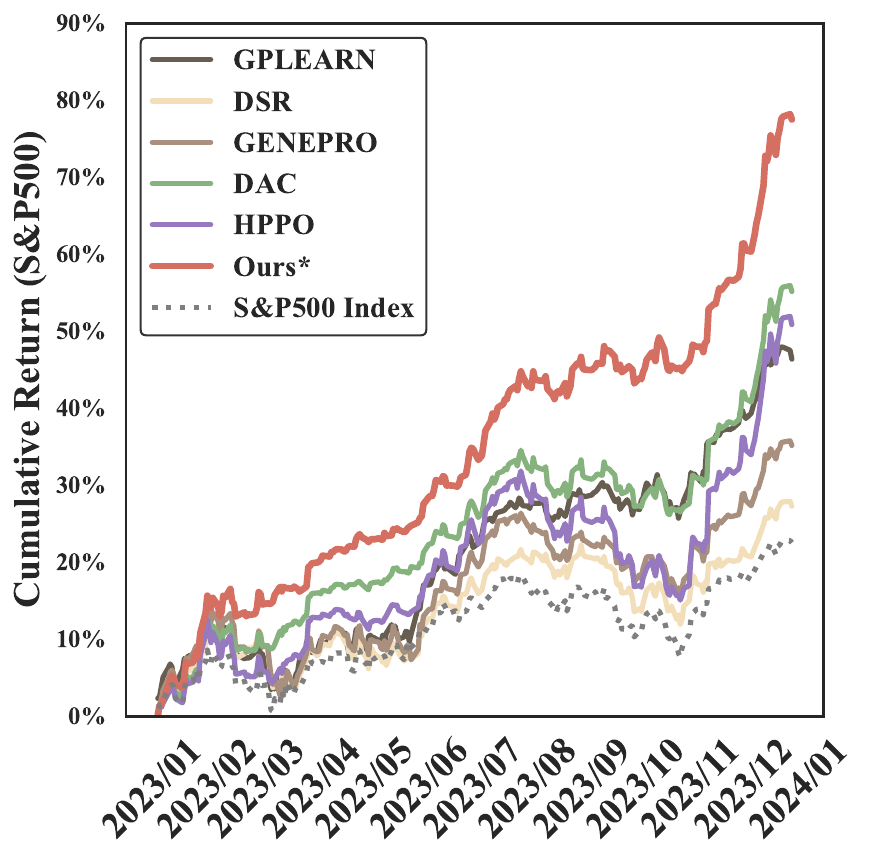}
    \caption{S\&P500 Index}
  \end{subfigure}%
  \hfill
  \begin{subfigure}{0.48\linewidth}
    \includegraphics[width=\linewidth]{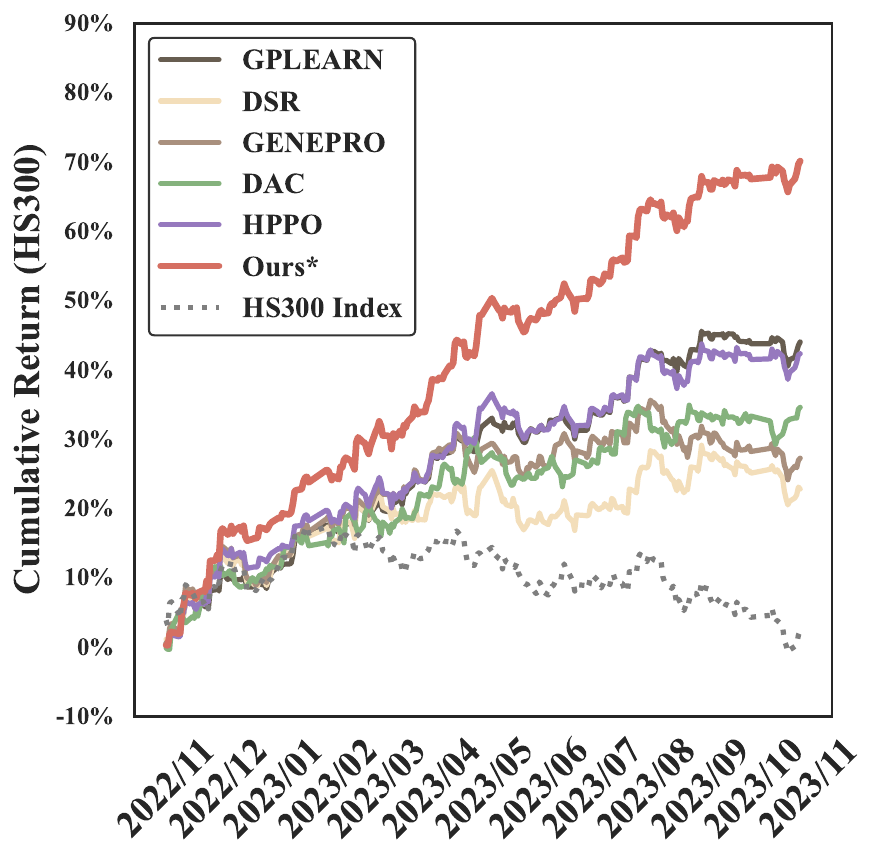}
    \caption{HS300 Index}
  \end{subfigure}
  \caption{Trading portfolio simulations: a backtesting comparison across different indexes.}
  \label{fig:2}
\end{figure}

\subsection{Main Results}
\par\noindent
\textbf{Comparison across all risk factor generators.} \quad
Experiments (Table~\ref{tab:comparison_methods}) on the HS300 and S\&P500 stock markets compare HRL-based (HPPO, DAC), DL-based (DSR, HRFT), and GP-based (GPLEA \\ RN, GENEPRO) risk factor generation methods. HPPO-TO consistently achieves the highest Normal ${IC}^*$, Rank ${IC}^*$, and ${IR}^*$ across both markets, outperforming all baselines. Among DL-based models, HRFT surpasses DSR in every metric, especially on HS300, where it posts the top ${IR}^*$ (0.6460) and competitive correlation scores, second only to HPPO-TO. In S\&P500, HRFT again outperforms DSR, demonstrating superior robustness. DSR is overall the weakest performer, often trapped in local optima due to its reliance on gradient descent. GP-based methods excel at global search; GPLEARN achieves higher Normal ${IC}^*$ and Rank ${IC}^*$ but trails GENEPRO in ${IR}^*$. GPLEARN outperforms DAC, as GP more effectively explores large solution spaces, while DAC is prone to premature convergence and higher computational costs. HPPO-TO outpaces GP methods by leveraging hierarchical exploration and transfer learning, which accelerate convergence and support robust subtask reuse. As a result, HPPO-TO delivers the strongest and most transferable correlations between generated risk factors and the target.

\par\noindent
\textbf{Comparison with varying factor pool capacities.} \quad
We further assess HPPO-TO’s performance with different factor pool sizes (\{10, 30, 50, 70, 90\}). Results in Figure~\ref{fig:1} confirm that HPPO-TO consistently outperforms all benchmarks across various HFT markets, with all methods improving as the pool size increases. GENEPRO generates factors with the lowest ${IR}^*$ but achieves higher Normal ${IC}^*$ and Rank ${IC}^*$ scores. HPPO ranks second, outperforming other baselines yet still trailing HPPO-TO.

\par\noindent
As shown in Figure~\ref{fig:2}, HPPO-TO scales effectively with the size of the risk factor pool and excels at identifying new risk factors. Its performance, as measured by ${IC}^*$ and ${\text{Rank}\ IC}^*$, surpasses all other methods in both the China (HS300) and U.S. (S\&P500) markets. Factors generated by HPPO-TO, GPLEARN, and HPPO display stronger target correlations. HPPO-TO and GPLEARN demonstrate greater prediction stability (${IR}^*$) than HPPO, while GENEPRO performs worst. HPPO-TO’s superior results stem from (1) decomposing risk factor construction into manageable subtasks and efficiently integrating learned skills, and (2) extracting and transferring common features across tasks by updating only the high-level policy, thereby simplifying the generation process.

\par\noindent
Table~\ref{tab:10hf_risk} lists five high-frequency risk factors generated by HPPO-TO for S\&P500 constituents with the highest ${IC}^*$ scores. These factors are randomly weighted across five weight sets. Notably, one factor exhibits an ${IC}^*$ exceeding 0.5, indicating a strong correlation with one-day-ahead RV. The top four factors are concise, with none exceeding a length of 15.

\subsection{Investment Simulation}
To evaluate the practical utility of risk factors, we implement a risk-averse portfolio strategy that selects the top 30 stocks based on factor values. We weight each stock using $w_i = \frac{1 / E_i(f)}{\sum_{j=1}^M \frac{1}{E_i(f)}}$ assigning lower weights to stocks with higher risk factors. Backtesting on HS300 and S\&P500 indices with 1/5-minute intraday data over one year (see Figure~\ref{fig:2}) shows HPPO-TO delivers the highest cumulative net value, outperforming HPPO by up to 25\%. All methods yield positive returns, with HPPO-TO, HPPO, and GPLEARN achieving substantial gains, while DAC lags. HPPO-TO and HPPO show similar performance patterns throughout the period.

\section{Conclusion}
In this study, we present a novel approach to automatically mine high-frequency risk factors, redefining traditional workflows in genetic programming-based risk factor extraction. Our proposed HPPO-TO algorithm, which integrates Hierarchical Reinforcement Learning (HRL) with transfer learning, achieves notable advancements in both the performance and efficiency of risk factor identification. Empirical results show that HPPO-TO has outperformed existing HRL and SR methods, achieving a 25\% excess investment return across major HFT markets, including China (HS300 Index) and the U.S. (S\&P500 Index).

\section{GenAI Usage Disclosure}
Generative AI (ChatGPT by OpenAI) was used solely for language editing and enhancing the readability of this manuscript. All scientific content, data analyses, results, and interpretations are entirely original and the sole work of the authors. No substantive content was generated or modified by AI tools.

\bibliographystyle{ACM-Reference-Format}
\bibliography{sample-base}


\end{document}